\documentclass[3p,times,procedia]{elsarticle}
\usepackage{nupha_ecrc}

\volume{00}

\firstpage{1}

\journalname{Nuclear Physics A}

\runauth{K.J. Eskola et al.}

\jid{nupha}

\jnltitlelogo{Nuclear Physics A}

\usepackage{amssymb}

\usepackage[figuresright]{rotating}

\begin{document}

\begin{frontmatter}



\dochead{XXVIth International Conference on Ultrarelativistic Nucleus-Nucleus Collisions\\ (Quark Matter 2017)}

\title{Latest results from the EbyE NLO EKRT model}


\author[label1,label2]{K. J. Eskola}
\author[label3]{H. Niemi}
\author[label1]{R. Paatelainen}
\author[label4,label2]{K. Tuominen}

 \address[label1]{University of Jyvaskyla, Department of Physics, P.O.B. 35, FI-40014 University of Jyvaskyla, Finland}
 \address[label2]{Helsinki Institute of Physics, P.O.B. 64, FI-00014 University of Helsinki, Finland}
 \address[label3]{Institut f\"ur Theoretische Physik, Johann Wolfgang Goethe-Universit\"at, Max-von-Laue-Str. 1, D-60438 Frankfurt am Main, Germany}
\address[label4]{Department of Physics, P.O.B. 64, FI-00014 University of Helsinki, Finland}



\begin{abstract}
We review the results from the event-by-event next-to-leading order perturbative QCD + saturation + viscous hydrodynamics (EbyE NLO EKRT) model.  With a simultaneous analysis of LHC and RHIC bulk observables we systematically constrain the QCD matter shear viscosity-to-entropy ratio $\eta/s(T)$, and test the initial state computation. In particular, we study the centrality dependences of hadronic multiplicities, $p_T$ spectra, flow coefficients, relative elliptic flow fluctuations, and various flow-correlations in 2.76 and 5.02 TeV Pb+Pb collisions at the LHC and 200 GeV Au+Au collisions at RHIC. Overall, our results match remarkably well with the LHC and RHIC measurements, and predictions for the 5.02 TeV LHC run are in an excellent agreement with the data. We probe the applicability of hydrodynamics via the average Knudsen numbers in the space-time evolution of the system and viscous corrections on the freeze-out surface.

\end{abstract}


\end{frontmatter}


\section{NLO EbyE EKRT model and its tests}
\label{Sec:intro}
The EKRT model \cite{Eskola:1999fc,Niemi:2015qia} rests on the idea that primary particle production in high energy heavy-ion collisions is dominated by few-GeV gluons, minijets \cite{Eskola:1988yh}, whose production rates are computable from collinear factorization of perturbative QCD (pQCD) but controlled by the phenomenon of saturation locally in the transverse plane \cite{Eskola:2000xq,Paatelainen:2012at,Paatelainen:2013eea}. The produced minijet densities can then be converted into initial conditions for relativistic fluid dynamics simulations.
In NLO pQCD, the infrared- and collinear-safe quantity computed here is the transverse energy $E_T$ carried by the minijets into a mid-rapidity window $\Delta y$ \cite{Eskola:2000ji_Eskola:2000my,Paatelainen:2012at} per transverse area $d^2\mathbf{r}$ in $A$+$A$ collisions at cms-energy $\sqrt{s_{NN}}$ and impact parameter $\mathbf{b}$,
\begin{equation}
\frac{dE_T}{d^2{\bf r}}(p_0, \sqrt{s_{NN}}, A, \Delta y, \mathbf{r}, \mathbf{b}; \beta) 
\stackrel{\rm pQCD}{=}
T_A(\mathbf{r}+ \mathbf{b}/2)T_A(\mathbf{r}- \mathbf{b}/2)\sigma\langle E_T \rangle_{p_0,\Delta y,\beta}
\stackrel{\rm saturation}{=}
\frac{K_{\rm sat}}{\pi}p_0^3\Delta y,
\label{eq:dET}
\end{equation}
where the transverse momentum cut-off $p_0\sim$ few GeV, and $T_A$ is the nuclear thickness function. The NLO quantity $\sigma\langle E_T \rangle_{p_0,\Delta y,\beta}$ is computed using collinear factorization and the subtraction method \cite{Kunszt:1992tn}. It contains the CTEQ6M parton distributions \cite{Pumplin:2002vw} with EPS09s nuclear effects \cite{Helenius:2012wd}, $2\rightarrow 3$ and UV-renormalized $2\rightarrow 2$ parton scattering matrix elements \cite{Ellis:1985er_Paatelainen:2014fsa}, and the measurement functions to define the $E_T$. The minimum $E_T$ in $\Delta y$ is controlled by the parameter $\beta \in [0,1]$, fixed to 0.8 here \cite{Paatelainen:2012at}. 
Saturation here is the limit where $E_T$ production from $(n>2)\rightarrow 2$ parton processes starts to dominate over the usual $2\rightarrow 2$ ones. This can be cast into the form of the saturation condition appearing on the r.h.s. of Eq.~(\ref{eq:dET}),
where $K_{\rm sat}$ is a free parameter \cite{Paatelainen:2012at}. Equation (\ref{eq:dET}) gives the saturation momentum
$p_0 = p_{\rm sat}(\sqrt{s_{NN}},A,\mathbf{r},\mathbf{b};\beta,K_{\rm sat})$
locally in the transverse plane. 
With a formation time $\tau_s(\mathbf{r}) = p_{\rm sat}(\mathbf{r})^{-1}$ the initial local energy density is then
\begin{equation}
e(\mathbf{r},\tau_{\mathrm{s}}(\mathbf{r})) = \frac{\mathrm{d}E_T}{\mathrm{d}^2\mathbf{r}}\frac{1}{\tau_{\mathrm{s}} (\mathbf{r}) \Delta y } = \frac{K_{\rm sat}}{\pi}[p_{\rm sat}(\mathbf{r})]^4.
\end{equation}

The key observation \cite{Paatelainen:2013eea,Eskola:2001rx} enabling the recently developed NLO EbyE EKRT model framework  of Ref.~\cite{Niemi:2015qia} is that 
$p_{\rm sat}(\mathbf{r},\mathbf{b})\approx p_{\rm sat}(T_AT_A)$ which can be parametrized. Then the $T_A$s can be made to fluctuate EbyE: we sample the nucleon positions from the standard Woods-Saxon density, setting a Gaussian gluon thickness function of a width $\sigma = 0.43$ fm \cite{Chekanov:2004mw} around each nucleon, and then computing the $T_A$ as a sum of these gluon clouds. Thus, the fluctuations of $T_A$ determine how
$e(\mathbf{r},\tau_{\mathrm{s}}(\mathbf{r}))$ fluctuates here EbyE. Finally, to start our hydro simulations at a constant time, we evolve the $e$-profile from $\tau_s(\mathbf{r})$ to $\tau_0 =1/p_{\rm sat}^{\rm min}=0.2$~fm using 0+1 D Bjorken hydrodynamics. At the edges of the system, we assume a binary $e$-profile.

\begin{figure}[hbt]
\begin{center}

\includegraphics[width=5.1cm]{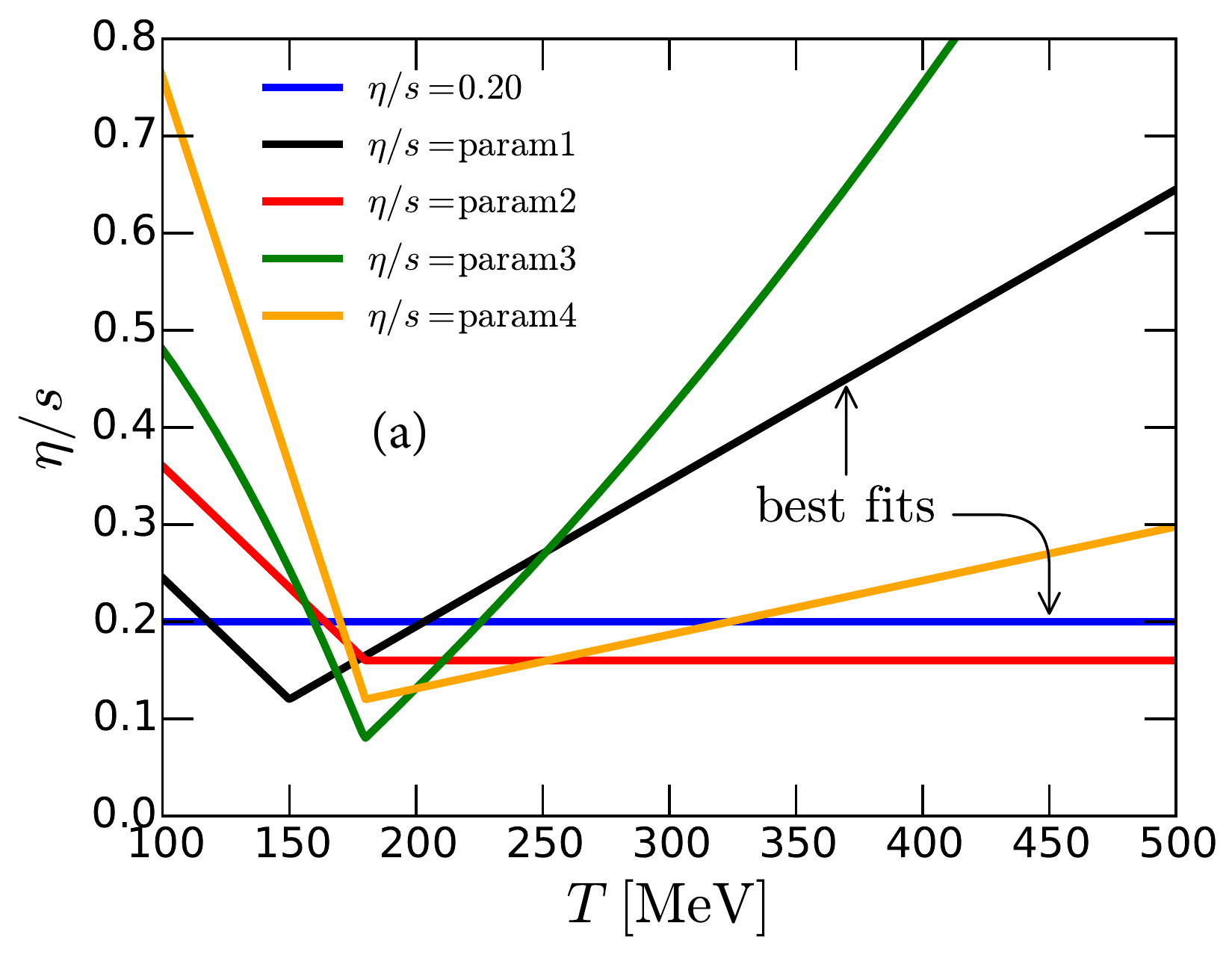}
\includegraphics[width=4.9cm]{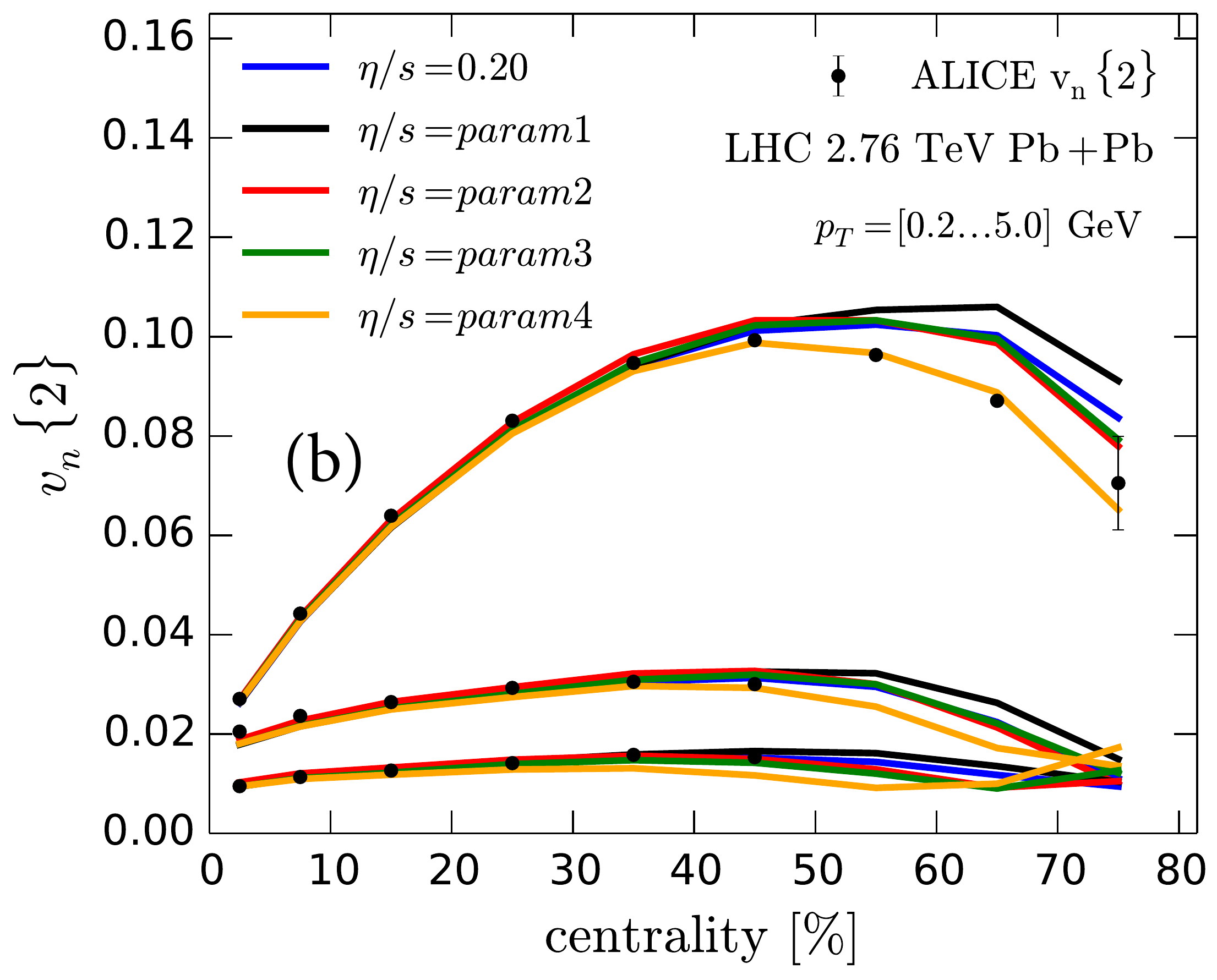}
\includegraphics[width=4.9cm]{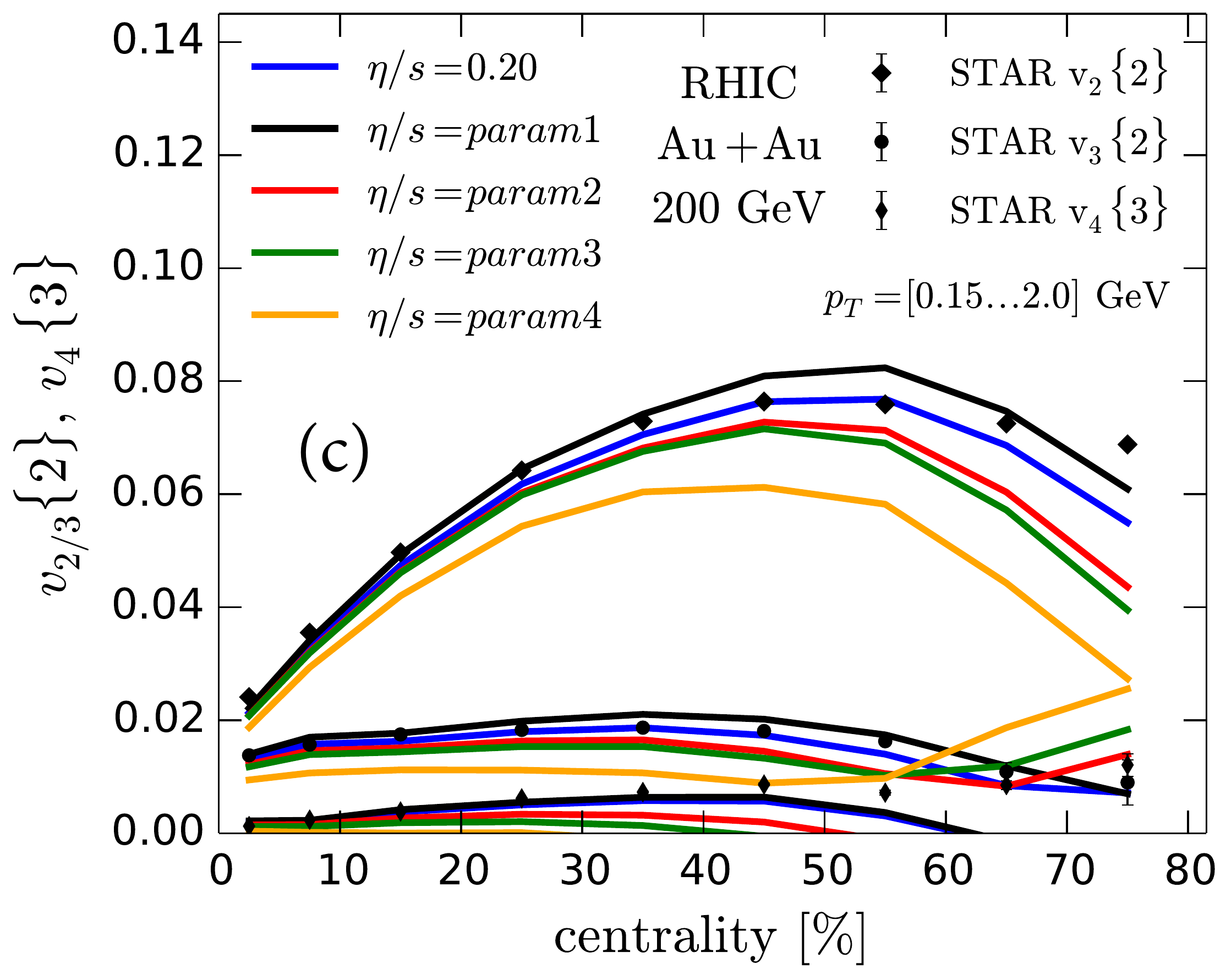}

\end{center}
\vspace{-0.5cm}
\caption{(a) The tested $\eta/s(T)$ parametrizations. 
Flow coefficients $v_n\{2\}$ vs. ALICE data \cite{ALICE:2011ab}
in 2.76 TeV Pb+Pb collisions at the LHC (b), and $v_{2}\{2\}$, $v_{3}\{2\}$ and $v_{4}\{3\}$ vs. STAR data \cite{Adams:2004bi, Adamczyk:2013waa, Adams:2003zg} in 200 GeV Au+Au collisions (c). From \cite{Niemi:2015voa,Niemi:2015qia}.}
\vspace{-0.1cm}
\label{fig:etapers}
\end{figure}

With such initial conditions, we describe the spacetime evolution of produced QCD matter then EbyE, using 2nd-order dissipative relativistic 2+1 D hydro with transient fluid-dynamics equation of motion for the shear-stress tensor $\pi^{\mu\nu}$ from Refs.~\cite{Denicol:2012cn, Molnar:2013lta}. The transverse flow and $\pi^{\mu\nu}$ are initially zero. Our  equation of state is $s95p$-PCE-v1 \cite{Huovinen:2009yb}, with chemical decoupling at $T_{\rm chem} = 175$ MeV.  Kinetic freeze-out is at $T_{\rm dec}=100$ MeV, and on this surface we assume, as usual, that the viscous $\delta f$-corrections
are  $\propto p_\mu p_\nu \pi^{\mu\nu}$. We neglect the bulk viscosity and heat conductivity. We study the $T$ dependence of $\eta/s(T)$ with the parametrizations of Fig.~\ref{fig:etapers}a,  all of which are designed to reproduce the flow coefficients $v_n\{2\}$ measured in 2.76 TeV Pb+Pb collisions at the LHC, as shown in Fig.~\ref{fig:etapers}b. The parameter $K_{\rm sat}$ is fixed separately for each  $\eta/s(T)$ parametrization, by using the $dN_{\rm ch}/d\eta(0-5 \%)$ measured by ALICE in 2.76 TeV Pb+Pb collisions (Fig.~\ref{fig:predictions}a).

We have extensively tested the NLO EbyE EKRT model in \cite{Niemi:2015qia}, arriving at a very good simultaneous description of the centrality dependences of charged hadron multiplicities, $p_T$ spectra, and flow coefficients in 2.76 TeV Pb+Pb collisions at the LHC and 200 GeV Au+Au at RHIC. As seen in Fig.~\ref{fig:etapers}c, the RHIC $v_n$s favor 0.2 (blue) and \textit{param1} (black) for $\eta/s(T)$. Also the correlations of 2 and 3 event-plane angles measured by ATLAS systematically favor these two $\eta/s(T)$ parametrizations, see Fig.~\ref{fig:symm_cum}a \cite{Niemi:2015qia}. Furthermore, these constraints are obtained in the centrality region where the $\delta f$ effects remain small in these observables \cite{Niemi:2015qia}. Relative EbyE fluctuations of $v_2$ measured by ATLAS provide a stringent $\eta/s$-independent test for the computed initial states. The EKRT model passes also this test remarkably well, demonstrating the necessity of a hydro evolution in understanding the centrality systematics of this observable \cite{Niemi:2015qia}.

As a measure of our hydro validity, we plot in Fig.~\ref{fig:symm_cum}f also 
\textit{(i)} the average Knudsen numbers $\langle {\rm Kn}\rangle$, expansion rate ($\theta = \partial_\mu u^\mu$) per thermalization time ($\tau_\pi=5\eta/(e+p)$) averaged over entropy density throughout the evolution ($T>100$~MeV), and 
\textit{(ii)} the shear stress over pressure $\langle\sqrt{\pi_{\mu\nu}\pi^{\mu\nu}}/p\rangle$ averaged over the entropy flux through the freeze-out surface. This reflects the average $\delta f$ corrections in the end of the evolution.
The facts that these indicators increase towards peripheral collisions only gradually and that $\langle {\rm Kn}\rangle={\cal O}(1)$ speak for the hydro validity at least up to 50\% centralities. Towards peripheral collisions, $\langle {\rm Kn}\rangle$ increases due to the increasing relative weight of the early stages where $\langle {\rm Kn}\rangle$ is large (see the $T>180$~MeV curve).

\section{Further predictions from the EbyE NLO EKRT model}
\label{Sec:latest}

\begin{figure}
\begin{center}

\includegraphics[width=4.6cm]{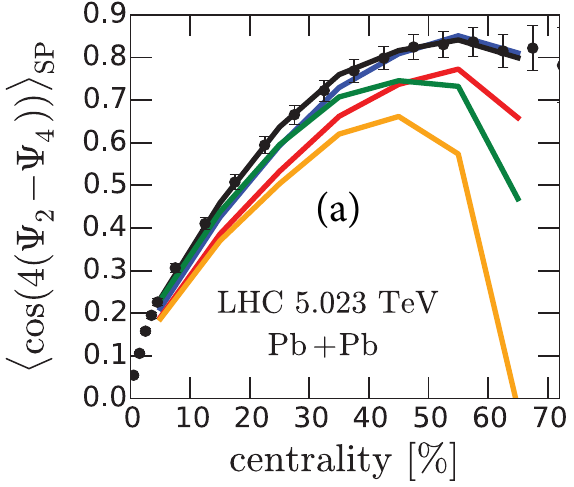}
\includegraphics[width=10.2cm]{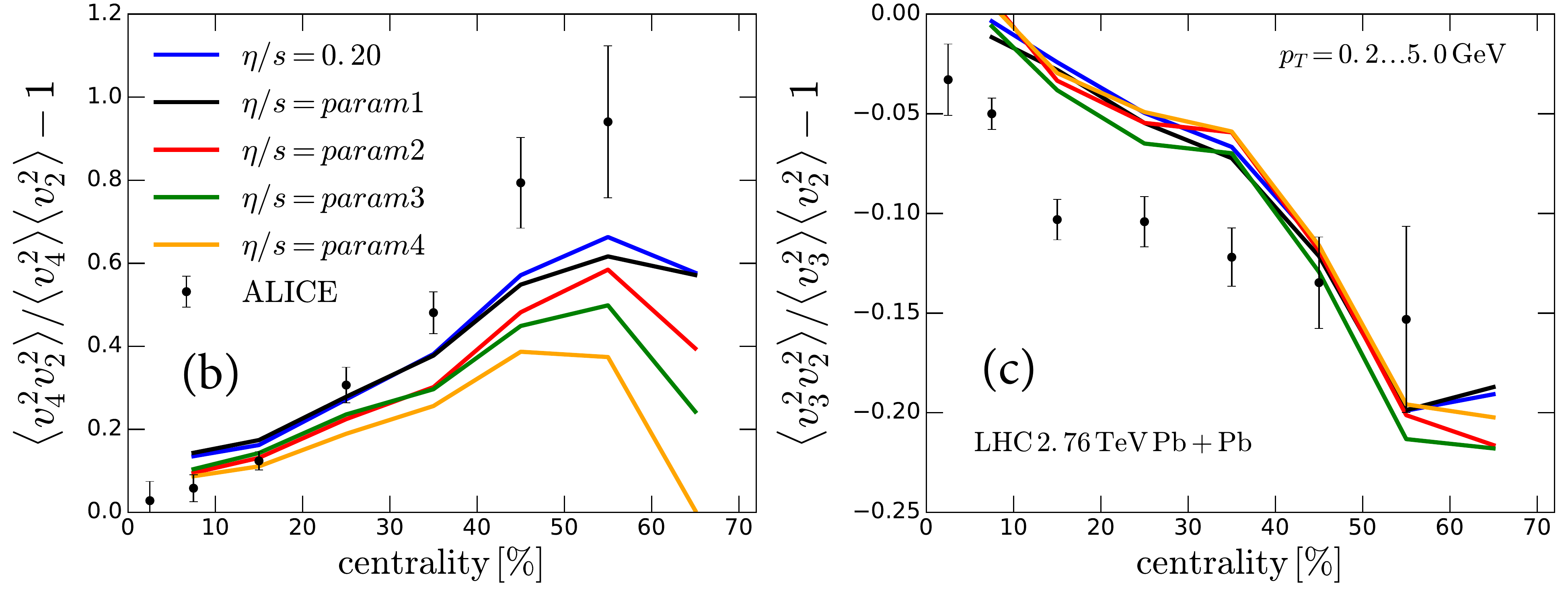}
\includegraphics[width=4.9cm]{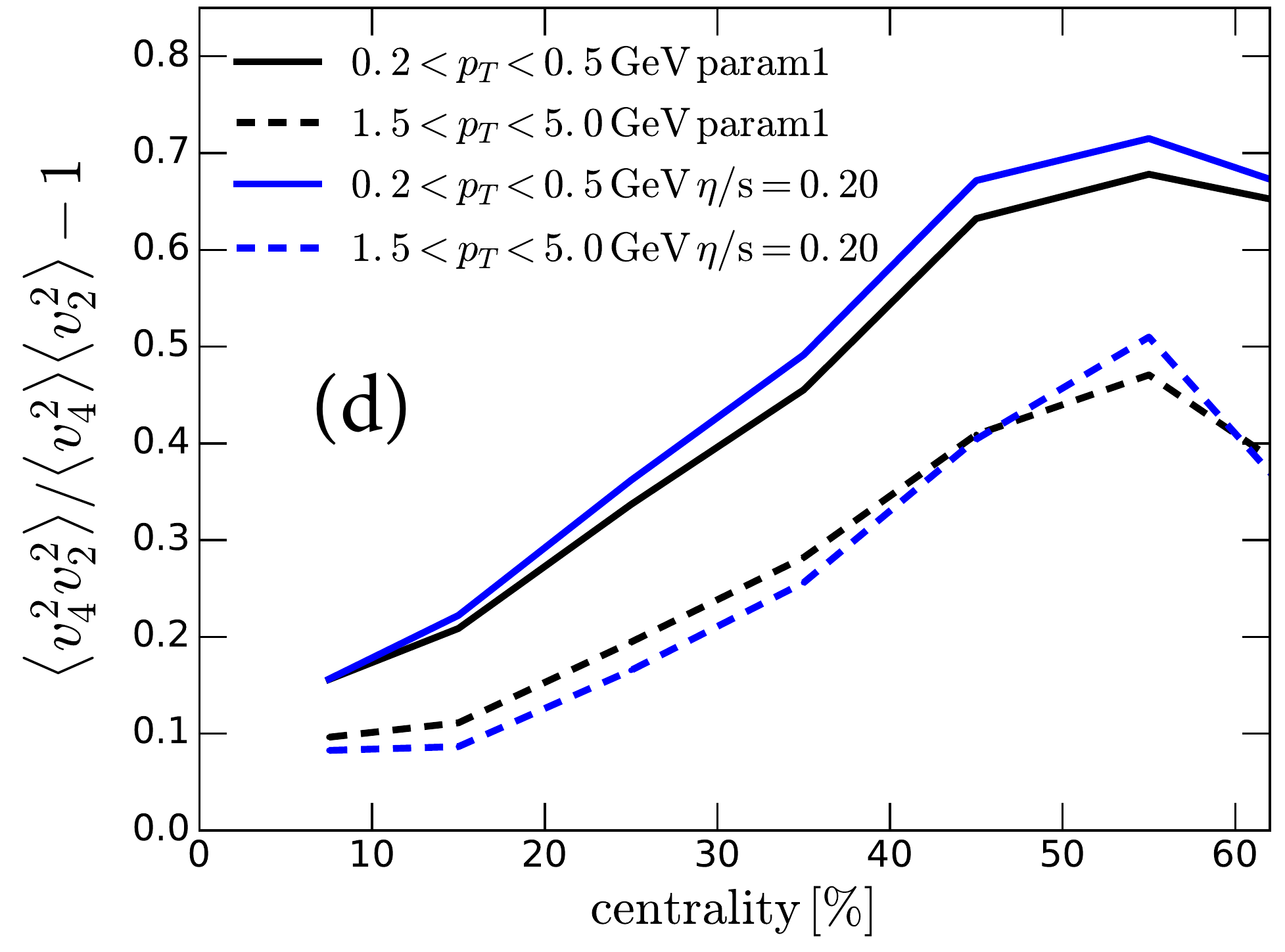}
\includegraphics[width=4.9cm]{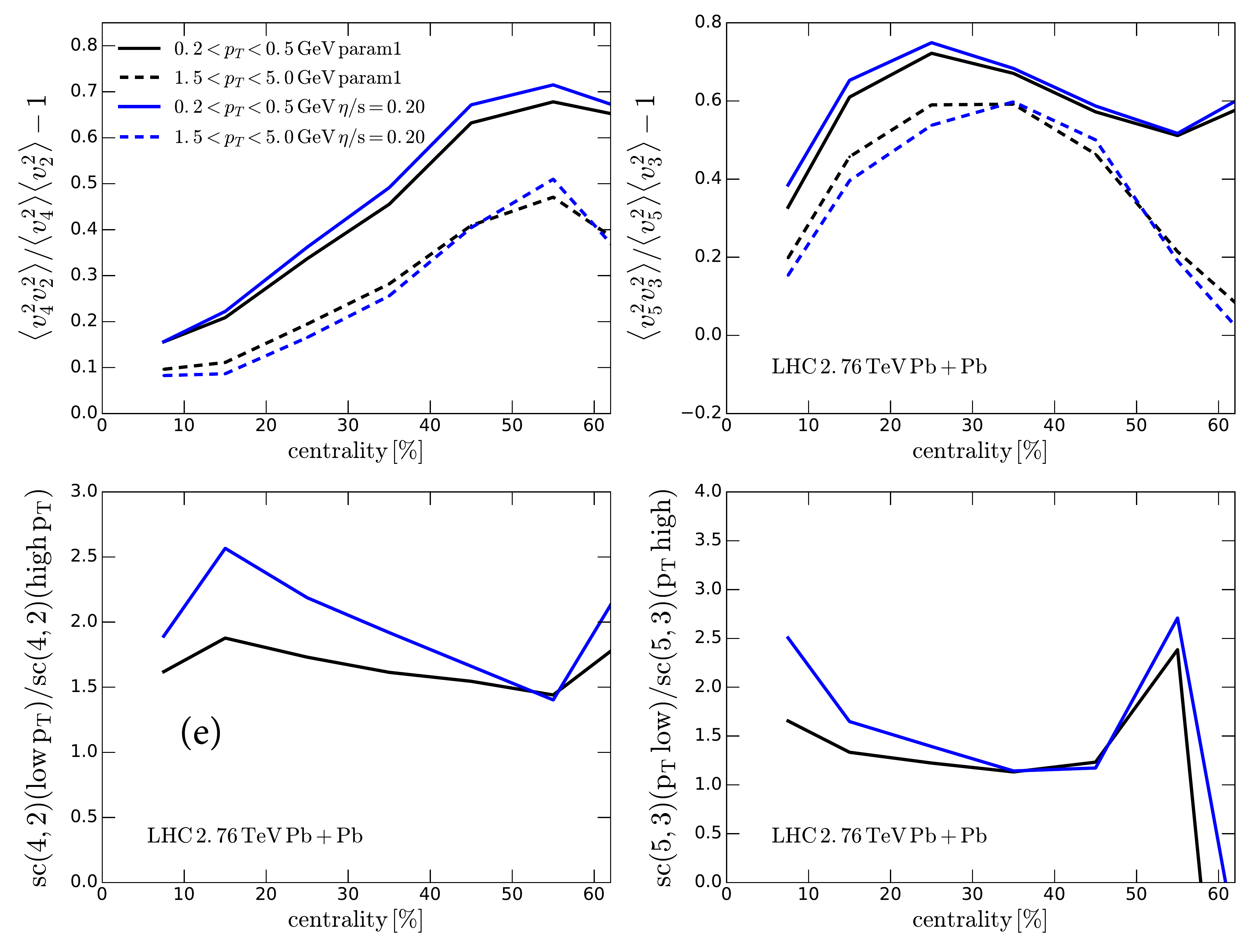}
\includegraphics[width=5.1cm]{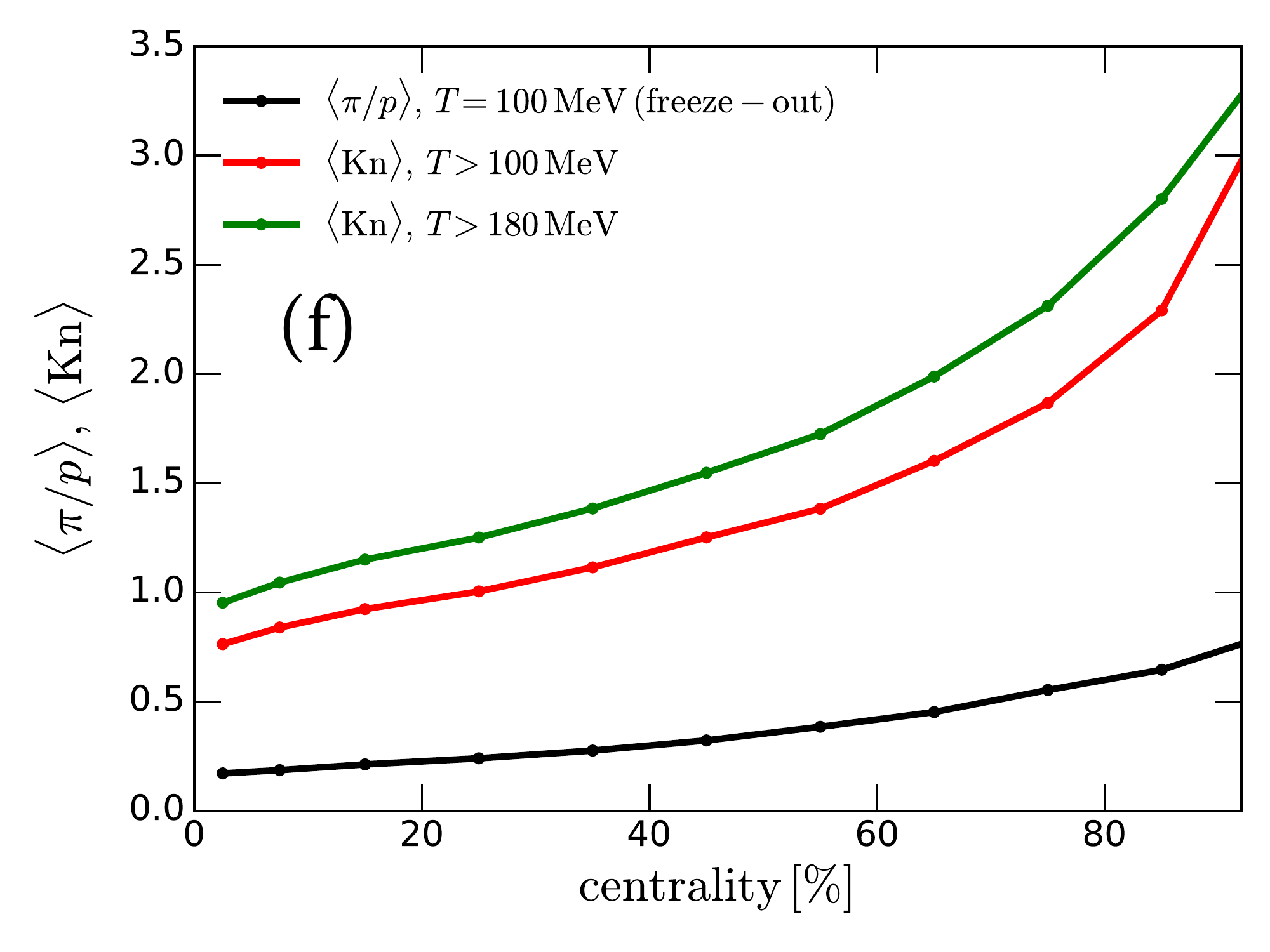}
\end{center}
\vspace{-0.5cm}
\caption{Centrality dependence of various correlators and Knudsen number in  2.76 TeV Pb+Pb collisions. 
(a) Correlation of the event-plane angles $\Psi_2$ and $\Psi_4$ vs. ATLAS data \cite{Aad:2014fla}. From \cite{Niemi:2015qia}.
(b) Normalized cumulants ${\rm SC}(4,2)/\langle v_4^2\rangle \langle v_2^2\rangle$ vs. ALICE data \cite{ALICE:2016kpq}.
(c) Same for ${\rm SC}(3,2)/\langle v_3^2\rangle \langle v_2^2\rangle$.
(d) ${\rm SC}(4,2)/\langle v_4^2\rangle \langle v_2^2\rangle$ in one low-$p_T$ and one high-$p_T$ interval, computed with our two best-fit $\eta/s$ parametrizations.
(e) Low-to-high-$p_T$ ratio of ${\rm SC}(4,2)/\langle v_4^2\rangle \langle v_2^2\rangle$.
(f) Average Knudsen numbers $\langle {\rm Kn} \rangle$ in our hydro evolution (red, green), and average shear stress over pressure $\langle \pi/p \rangle$ on the freeze-out surface (black), computed with the \textit{param1} $\eta/s$ parametrization.
}
\vspace{-0.3cm}
\label{fig:symm_cum}
\end{figure}

We have made a series of predictions from the EbyE NLO EKRT model without any further tuning. For ALICE, we have computed the symmetric 2-harmonic 4-particle cumulants, ${\rm SC}(m,n)=\langle\langle \cos(m\phi_1+n\phi_2-m\phi_3-n\phi_4)\rangle\rangle = \langle v_m^2v_n^2\rangle - \langle v_m^2\rangle \langle v_n^2\rangle$ normalized by $\langle v_m^2\rangle \langle v_n^2\rangle$ shown in Fig.~\ref{fig:symm_cum}b,c.
Our best-fit $\eta/s$ parametrizations predict rather well the positive correlation seen by ALICE \cite{ALICE:2016kpq} in ${\rm SC}(4,2)$ and also the trend of the negative correlation in ${\rm SC}(3,2)$. We emphasize, however, the importance of a 1-to-1 comparison: we expect that once we include the multiplicity weighting assumed in the ALICE analysis, our prediction will be systematically closer to the data.
In Fig.~\ref{fig:symm_cum}d we show a prediction of the $p_T$ dependence of  ${\rm SC}(4,2)/\langle v_4^2\rangle \langle v_2^2\rangle$. Fig.~\ref{fig:symm_cum}e in turn suggests that the low-to-high-$p_T$ ratios of these normalized correlators might be able to distinguish between our best-fit $\eta/s$ parametrizations.
Similarly, we have provided the STAR collaboration with our predictions for the centrality dependence of mixed harmonic correlators $C_{m,n,m+n}=\langle\langle\cos(m\phi_1+n\phi_2-(m+n)\phi_3)\rangle\rangle$.
As shown in \cite{Adamczyk:2017byf}, our best-fit parametrizations reproduce the $C_{2,2,4}$ rather well. However, we underestimate the measured $C_{2,3,5}$, which we believe is due to large $\delta f$ effects in this observable, possibly combined also with non-flow and rapidity effects which we cannot consider, yet. Further studies on this are ongoing.

\begin{figure}
\begin{center}
\includegraphics[width=6.1cm]{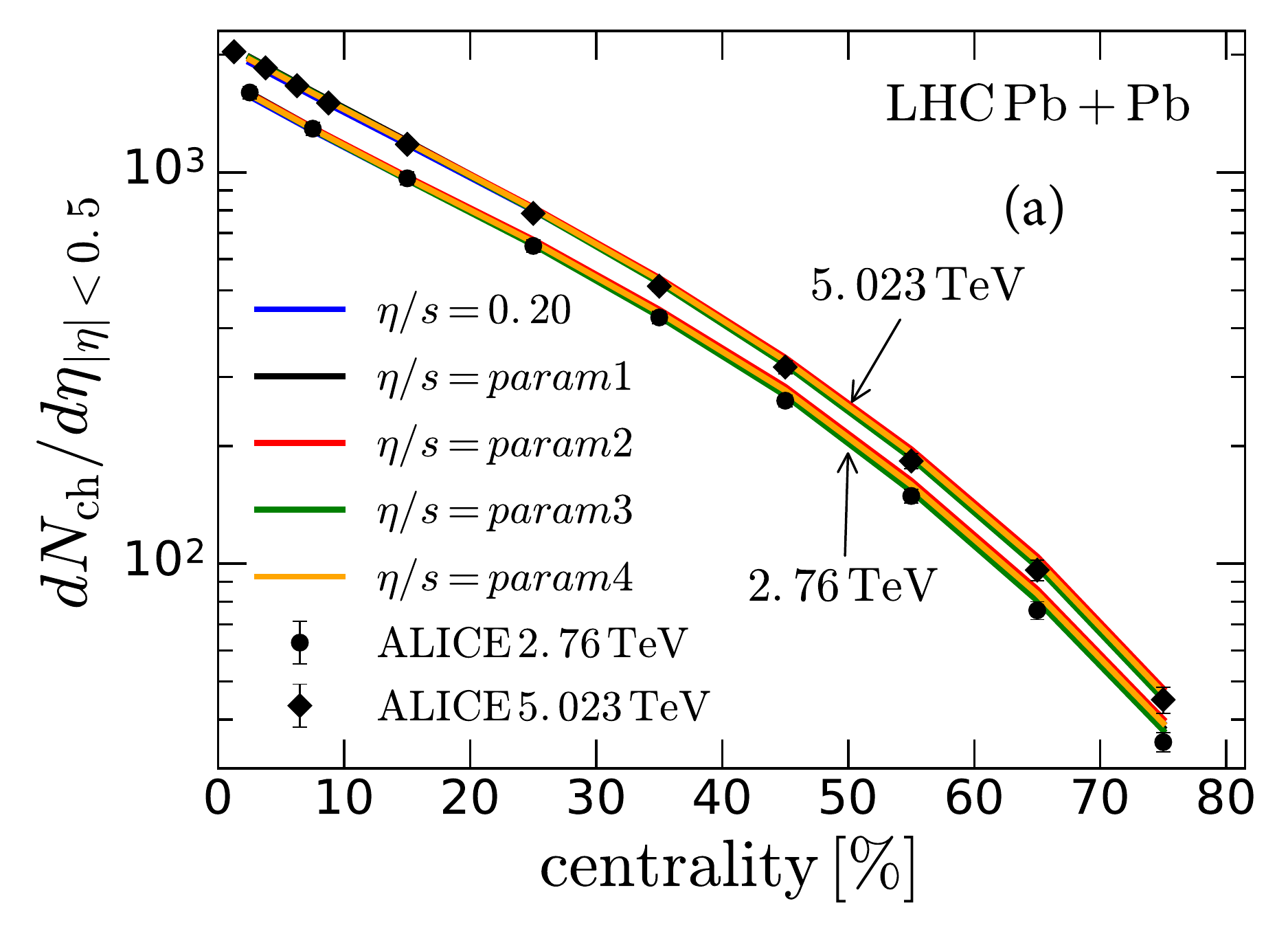}
\includegraphics[width=6.1cm]{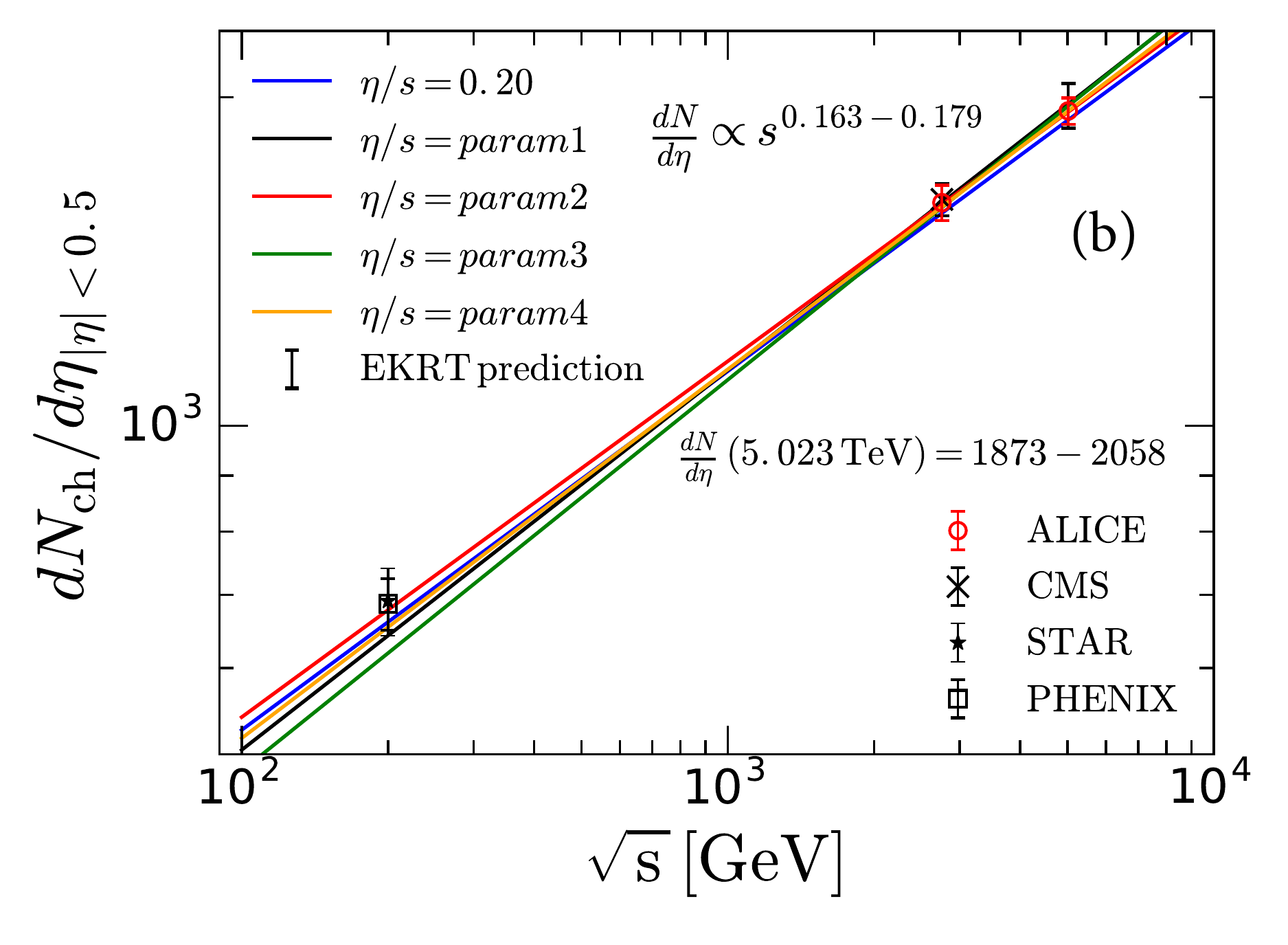}

\vspace{-0.2cm}

\includegraphics[width=10.5cm]{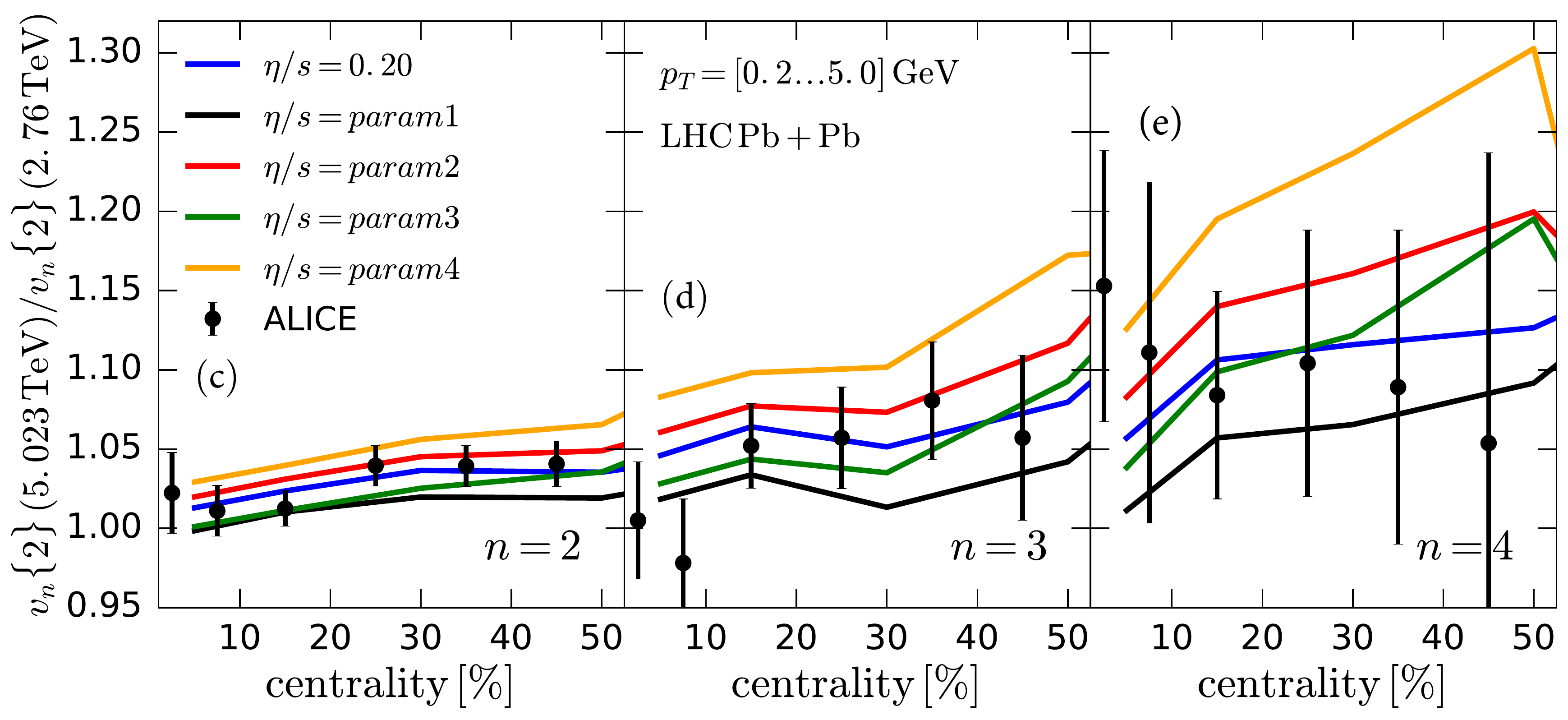}
\end{center}
\vspace{-0.5cm}
\caption{EbyE NLO EKRT model predictions for 5.023 TeV Pb+Pb collisions \cite{Niemi:2015voa}. 
(a) Centrality dependence of charged particle multiplicity, vs. ALICE data \cite{Aamodt:2010cz,Adam:2015ptt}. 
(b) Predicted $\sqrt{s_{NN}}$ dependence of charged particle multiplicity from RHIC Au+Au to LHC Pb+Pb collisions vs. data from ALICE \cite{Aamodt:2010cz,Adam:2015ptt}, CMS \cite{Chatrchyan:2011pb}, STAR \cite{Abelev:2008ab} and PHENIX \cite{Adler:2004zn}.  
(c-e) Ratio of the flow coefficients $v_n\{2\}$ in 5.023 TeV and 2.76 TeV Pb+Pb collisions, vs. ALICE data \cite{Adam:2016izf}.
}
\vspace{-0.5cm}
\label{fig:predictions}
\end{figure}

Thanks to the predictive power of the EKRT model, we have also made predictions for the 5.02 TeV Pb+Pb run at the LHC \cite{Niemi:2015voa}. Figure \ref{fig:predictions}
shows our predictions for the multiplicity and flow-coefficient ratios. In the latter, notice the slight increase with increasing $n$.
Again, as seen in the figure, the EbyE NLO EKRT model fairs very well in the data comparison.

To conclude, the EbyE NLO EKRT model \cite{Niemi:2015qia} explains consistently the bulk observables and various correlators at mid-rapidity in LHC and RHIC heavy-ion collisions. Its predictive power in cms-energy, centrality and nuclear mass number has been demonstrated with various observables. Via a multi-energy and multi-observable analysis we have managed to constrain the $\eta/s(T)$ ratio, for which two best-fit parametrizations have been identified. Similar results have been found also in Ref.~\cite{Bernhard:2016tnd}.
Systematic further tests of the hydro results validity are, however still needed, especially in the case of more complicated correlators, as well as more work for including further dissipative phenomena. 

\vspace{0.1cm}

\noindent{\small\textbf{Acknowledgments}.\ 
K.J.E.\ is supported by the Academy of Finland, Project 297058, and
H.N.\ by the EU's Horizon 2020 research and innovation programme under the Marie Sklodowska-Curie grant agreement no.\ 655285.}








\end{document}